\def   \ni {\noindent}

\def   \ssk {\vskip  5truept}

\def   \bsk {\vskip 15truept}

\def   \newline {\hfil\break}

\def\ol#1{\overline{#1}}
\def\beq{\begin{equation}}
\def\eeq{\end{equation}}
\def\bea{\begin{eqnarray}}
\def\eea{\end{eqnarray}}

\documentstyle[epsfig]{article}
\begin{document}

%
\def\la{\mathrel{\mathchoice {\vcenter{\offinterlineskip\halign{\hfil
$\displaystyle##$\hfil\cr<\cr\sim\cr}}}
{\vcenter{\offinterlineskip\halign{\hfil$\textstyle##$\hfil\cr
<\cr\sim\cr}}}
{\vcenter{\offinterlineskip\halign{\hfil$\scriptstyle##$\hfil\cr
<\cr\sim\cr}}}
{\vcenter{\offinterlineskip\halign{\hfil$\scriptscriptstyle##$\hfil\cr
<\cr\sim\cr}}}}}
\def\ga{\mathrel{\mathchoice {\vcenter{\offinterlineskip\halign{\hfil
$\displaystyle##$\hfil\cr>\cr\sim\cr}}}
{\vcenter{\offinterlineskip\halign{\hfil$\textstyle##$\hfil\cr
>\cr\sim\cr}}}
{\vcenter{\offinterlineskip\halign{\hfil$\scriptstyle##$\hfil\cr
>\cr\sim\cr}}}
{\vcenter{\offinterlineskip\halign{\hfil$\scriptscriptstyle##$\hfil\cr
>\cr\sim\cr}}}}}
\def\degr{\hbox{$^\circ$}}
\def\arcmin{\hbox{$^\prime$}}
\def\arcsec{\hbox{$^{\prime\prime}$}}

\hsize 5truein
\vsize 8truein
\font\abstract=cmr8
\font\keywords=cmr8
\font\caption=cmr8
\font\references=cmr8
\font\text=cmr10
\font\affiliation=cmssi10
\font\author=cmss10
\font\mc=cmss8
\font\title=cmssbx10 scaled\magstep2
\font\alcit=cmti7 scaled\magstephalf
\font\alcin=cmr6 
\font\ita=cmti8
\font\mma=cmr8
\def\ref{\par\noindent\hangindent 15pt}
\null


\title{\ni Detecting large-scale deviations from FRW geometry 
     with future CMB measurements}

\bsk \bsk
\author{\ni G.~Dautcourt}

\bsk
\affiliation{\ni Max-Planck Institut f\"ur Gravitationsphysik, 
Albert-Einstein-Institut \\
Haus 5, Am M\"uhlenberg, D~--~14476 Golm, Germany}
\bsk
\baselineskip = 12pt

\abstract{{\bf{\scriptsize{ABSTRACT}}}

\ni 
We discuss the question to what degree the geometrical structure
and the matter content of the universe at scales exceeding the 
present Hubble horizon is constrained by cosmological observations, 
in particular by measurements of the cosmic microwave background
radiation. For an answer, a simple formalism is described, 
which goes back to a paper by Kristian and Sachs in 1966.}   
\bsk
\baselineskip = 12pt
\bsk
\baselineskip = 12pt


\text{\ni {\bf{1. INTRODUCTION}}
\ssk
\ni

It is usually assumed that the observable universe can be
described by one of the homogeneous-isotropic
Friedman-Robertson-Walker models,
at least up to scales of the order of the horizon $\lambda_H$.
The structure of the universe at larger scales
is not known. The inflationary scenario suggests, that the
observable universe is part of an inflated region of size 
$\lambda_I$, possibly much larger than $\lambda_H$.
Outside the inflated and hence nearly homogeneous region
the spacetime geometry might be inhomogeneous and anisotropic.
Could structures outside the horizon by means of their tidal fields
have some imprint on the observationally accessible part of 
the universe (shortly: the local universe), in particular, 
may they influence the cosmic microwave background radiation?

Twenty years ago L. Grishchuk and Ya. B. Zeldovich (1978) (see 
also Turner 1991, Grishchuk 1992, Kashlinsky et al. 1994,
Lyth 1995, Garcia-Bellido  et al. 1995) have
discussed a similar question. They found that the CMB perturbations
$\delta T/T$ resulting from the growing mode of density 
or gravitational wave perturbations on a scale $\lambda$
and corresponding to a metric perturbation $h$ are of the order 
$\delta T/T \sim h (\lambda_H/\lambda)^2$. Since one expects 
$h\sim 1$ for regions $\lambda \ge \lambda_I$, the COBE 
observations lead to $\lambda_I > 10^3\lambda_H$, thus the 
"homogeneous region" must be much larger than the Hubble distance,
independent of any inflationary hypothesis. {\it This argument 
does not exclude structures exceeding $\lambda_H$ only slightly, 
if they correspond to metric amplitudes $h$ much smaller than 1}. 
Also other deviations from FRW models may be allowed to a certain
degree. For instance, if one restricts the geometry to homogeneous 
Bianchi models, one is able to obtain from the CMB measurements 
\eject
\ni
upper limits for the vorticity and shear of these models (Hawking 
1969, Collins and 
Hawking 1973, Bunn et al. 1996, Kogut et al. 1997). 
But constraining 
the geometry to a small class of simple models may be misleading, 
since the real universe might have a much more complicated 
geometrical structure and matter content at very large distances.
In view of the remarkable improvements expected for future CMB 
observations with the MAP and PLANCK missions it is of interest to 
see {\it how much of this structure could be detected or 
constrained}. A first step is a mathematical framework dealing 
with sufficiently general cosmological models. We describe a very
simple proposal, which tackles the problem, but needs further 
elaboration.

\bsk
\ni {\bf{3. KRISTIAN-SACHS APPROACH}}
\ssk
\ni 
J.~Kristian and R.K.~Sachs (1966) 
made in their classical paper only few assumptions:
(i) The universe is described by a Riemannian spacetime with slowly
     varying metric tensor;
(ii) light travels along null geodesics and obeys the usual 
     area-intensity law;
(iii) the gravitational field is related to the matter by the 
     Einstein field equation for dust;
(iv) all quantities of interest can be expanded in a power series 
     around here-and-now as origin.
In their treatment it not necessary to adopt a specific
cosmological model a priori, the results are complete and general 
(apart from the assumption of dust, which can quickly be
generalized). Ignoring global properties (or rather assuming that 
the scale of a possible compactification of the universe is larger 
than all other scales) made the calculation manageable, and last 
not least all introduced quantities are measurable in principle.

With a power series expansion around here-and-now one can hardly  
treat evolutionary problems in cosmology. In particular, 
the CMB fits not easily into the framework.
We remove this shortcoming by treating space and time
coordinates on a different footing. Temporal variations are 
seen to appear with much higher amplitude than spatial
variations - at least if one averages over the inevitable 
small-scale fluctuations as origin of galactic structures.
This leads to the idea to carry out the power-law expansion
only within spatial hyper-surfaces orthogonally intersecting the
observers world line. The Kristian-Sachs expansion coefficients
then become time-dependent.
Actually, we reduce the generality by adopting the 
(anthropomorphic) assumption, 
that space-time reduces along the observers world line to the 
homogeneous-isotropic FRW models. In geometrical terms,
the observers world line $L$ is assumed as geodesic and shear-free,
generally in contrast to the world-lines of neighbouring observers.
The method remembers to some degree the covariant approach put 
forward by G. Ellis and his coworkers in numerous publications, 
but contrary to their work we try to fix the coordinate system as 
much as possible in order to reduce the number of coefficients 
needed to describe deviations from FRW models. 

\bsk
\ni {\bf{3. COORDINATE SYSTEM }}
\ssk
\ni 

Basic to the method is the construction of a coordinate system,
which is adapted to the matter distribution. We represent the 
cosmic matter as relativistic fluid flow with $V^\mu$ as the 
four-velocity vector tangent to the flow lines 
($V^\mu V_\mu = -1$, our notation follows the book by Misner, 
Thorne and Wheeler 1973).
Comoving coordinates are introduced such that $V^{\mu} = 
\delta^{\mu}_0$. The flow lines are given by 
$x^i= const ~(i=1,2,3)$. The time coordinate $t$ is not completely 
fixed by its coincidence with the proper time on the flow lines, 
and also the spatial coordinates are subject to arbitrary 
time-independent changes, 
$\ol t = t+f(x^i),~ \ol x^i = \ol x^i(x^k)$. 

Let the particular flow line $x^i=0$ be the observers world 
line $L$. We demand that on and near $L$ a FRW solution (with 
zero spatial curvature, see below) is a good approximation. 
It is sufficient to assume that $L$ (and only this world line) is 
geodesic and shear-free. This is a weak form of reducing to FRW 
along $L$, since it is compatible with a non-vanishing Weyl tensor 
and a nonzero vorticity on $L$. We may then represent the metric 
tensor near $L$ as a Taylor series in the spatial coordinates $x^i$:
\bea g_{00} &=& -1  \\
 g_{0i} &=& l_{ik}x^k + l_{ikl}x^kx^l + l_{iklm}x^kx^lx^m + ...\\
 g_{ik} &=& a^2\delta_{ik}+h_{ikl}x^l + h_{iklm}x^lx^m 
+ h_{iklmn}x^lx^mx^n + ... 
\eea           
where $a(t)$ is the Friedman scale factor, the other expansion
coefficients are as well time functions in general.  
The coordinate transformations preserving this form of the metric 
may also be expanded with constant coefficients, which later serve 
to reduce initial values of the metric expansion coefficient:
\bea
\ol t &=& t +a_{kl}x^kx^l +a_{klm}x^kx^lx^m + ..., \\  
\ol x^i & =& x^i + b_{ikl}x^kx^l + b_{iklm}x^kx^lx^m + ... . 
\eea 
Only in a linear approximation the terms added to the FRW metric 
can uniquely be separated into scalar, vector and tensor 
(gravitational wave) perturbations, we shall not attempt this here.
 - We have assumed a {\it flat} FRW metric along $L$, since the 
spatial curvature of the FRW models is hidden in the expansion 
coefficients of Eqn (3).

\bsk
\ni {\bf{4. MATTER CONGRUENCE }}
\ssk
\ni 

Using the decomposition of the velocity gradient
\beq V_{\mu;\nu} = \omega_{\mu\nu} +\sigma_{\mu\nu} 
+\Theta(g_{\mu\nu}+V_{\mu}V_{\nu})/3 - V_{\nu}A_{\mu},
\eeq 
we can expand acceleration, shear, expansion and vorticity around 
$L$:
\bea  A_i &=& (\dot{l}_{ik} + \dot{l}_{ikl}x^l)x^k + o(3),\\
\sigma_{kl} &=& \frac{1}{2}(\dot{h}_{klr}
- \frac{1}{3}\delta_{kl}\dot{h}_{ssr})x^r +o(2),\\
\Theta &=& 3\frac{\dot{a}}{a} 
+\frac{1}{2}\frac{d}{dt}(\frac{h_{kkl}}{a^2})x^l 
 + o(2), \\
\omega_{kl} &=& \frac{1}{2}(l_{kl}-l_{lk}) 
+(l_{klr}-l_{lkr})x^r + o(2),
\eea
together with $A_0=0,~\sigma_{0\mu}=0,~\omega_{0k}=0$. 
The lowest terms in this expansion transform under (3),(4) as 
\bea \ol l_{kl}(\ol t) &=& l_{kl} + 2a_{kl}, \\
\ol h_{klm}(\ol t) &=& h_{klm}(t) - 2 a^2(b_{klm}+b_{lkm}).
\eea          
Since $a_{kl}=a_{lk}$, the vorticity $\omega_{kl}$ on $L$ is not
affected by coordinate transformations. Similar simple transformation
rules hold for the higher-order coefficients.
For the matter tensor we assume a perfect fluid
\begin{equation}T_{\mu\nu}= V_{\mu}V_{\nu}(\mu + p)+pg_{\mu\nu}
\end{equation}
with an equation of state $p=w\mu~ (w=const)$.
The conservation equations $T^{\mu\nu}_{;\nu}=0$ can partly be 
integrated without expanding $\mu$ and $p$ in a Taylor series. 
One obtains
\beq \mu = m(x^i)((1+g_{0k}g_{0l}\gamma^{kl})det(g_{ik}))^{-(1+w)/2},
\eeq
\beq
 0 = \dot g_{0k} 
+ \frac{w}{1+w}(\frac{\mu_{,k}}{\mu} +g_{0k}\frac{\dot\mu}{\mu}),
\eeq
$\gamma^{ik}$ is the inverse matrix to $g_{ik}$. We are here 
particularly interested in the case of dust $w=0$, which is an
approximation accurate enough to illustrate the method, when we 
integrate the null geodesics back to the surface of last scattering 
of the CMB photons. In the dust case the fluid is not accelerated, 
and $g_{0i}$ is time-independent. $g_{0i}$ is not a gradient, when 
vorticity (the antisymmetric part in $l_{ik}$) is present. We may 
however use Eqn (11) to transform the symmetric part of $l_{ik}$ 
to zero. For more information we have to look at the field equations. 

\bsk
\ni {\bf{5. FIELD EQUATIONS}}
\ssk
\ni 
The Einstein field equations cannot be integrated as easily as the 
matter conservation equations, so we use Taylor expansion again. 
It is straightforward to write them down for the metric (1)-(3) 
and the matter tensor (13). To zero order one obtains,
again assuming dust and with $\rho_0$ as matter density on $L$: 
\beq 3\frac{\stackrel{..}{a}}{a} +\frac{1}{2}\kappa\rho_0 
=\frac{\omega_{kl}\omega_{kl}}{a^4},
\eeq
\beq 0= \frac{1}{2} \frac{d}{dt}(\frac{h_{llk}-h_{kll}}{a^2})
+\frac{2}{a^2}(l_{llk}-l_{kll}) 
+ \frac{1}{a^4}\left(\omega_{lr}(h_{lkr}-h_{rkl})
+ \omega_{lk}(-2h_{lrr}+h_{rrl})\right)
\eeq
\bea  \delta_{kl}(\frac{\stackrel{..}{a}}{a} +2\frac{\dot a^2}{a^2}
 - \frac{1}{2}\kappa\rho_0) = && \nonumber \\
 \frac{1}{a^4}\omega_{kr}\omega_{lr}
+ \frac{1}{a^4}(h_{klrr}+h_{rrkl}-h_{rlkr}-h_{krlr}) &&\nonumber\\
+ \frac{1}{4a^6}(2h_{srr}-h_{rrs})(h_{skl}+h_{slk}-h_{kls})&&  
\nonumber\\
- \frac{1}{4a^6}(h_{srl}+h_{slr}-h_{rls})(h_{skr}+h_{srk}-h_{krs}). 
\eea
We have written the FRW terms on the lhs and the terms
representing external sources on the rhs. We have also
kept all nonlinear terms to see the structure of the equations,
even if these terms can be neglected for a first calculation of
the effects of external gravitational fields. The equations show 
that corrections to the Friedman equations occur already at zero 
order, i.e., on the observer world-line $L$. The time dependence of
the expansion coefficients is only partly determined by the field
equations at the given level, thus one has to take higher levels 
into account. Some equations become algebraic relations between the 
coefficients. We note 
that the restrictions for the expansion 
coefficients in (1)-(3) found in this way are only {\it necessary} 
conditions for $g_{\mu\nu}$ to represent a solution of the field 
equation. Our expansion procedure is as yet not based on a 
well-founded initial value problem, and it is therefore not 
sure that every approximate expressions for $g_{\mu\nu}$ 
can be extended to a full solution of the Einstein field equations. 

\bsk
\ni {\bf{6. THE CMB: ORIGIN OF A COSMOLOGICAL DIPOLE COMPONENT}}
\ssk
\ni
The CMB anisotropy in our local cosmological model  
can be calculated  using a well-known relation derived 
by Panek (1986), Russ et al. (1993) and Dunsby (1997). We write 
the equation as
\beq
\frac{\delta T}{T} = 
-\int^{t_0}_{t_1}\left(\frac{1}{4\mu_{(r)}p^0}
(p^{\rho}\mu_{(r),\rho} -\dot{\mu}_{(r)}E)
+\frac{\sigma_{\mu\nu}p^{\mu}p^{\nu}}{Ep^0} 
+\frac{\dot{V}_{\mu}p^{\mu}}{p^0} \right)dt, \label{dunsby}
\eeq
where $p^{\mu}= \frac{dx^{\mu}}{ds}$ is the tangential vector of a 
null geodesic connecting the emission (at time $t_1$) of the 
radiation with the recording event (at time $t_0$), $\mu_{(r)}$ 
the density of the radiation field and $E= -p^{\mu}V_{\mu}$ the 
photon energy relative to $V_{\mu}$. Using this relation requires 
integration of null-geodesics, i.e. solution of the geodesic equation 
$p^{\mu}_{;\mu}=0$. This can easily be done within the formalism, 
if one considers the additional external terms as perturbations 
to the known photon motion in a FRW geometry. The external or 
super-horizon modes under discussion affect the lowest-order 
multipoles. As discussed by Bunn et al. (1996), they must be 
distinguished from the familiar statistical   
fluctuations in $T$, which correspond to an isotropic Gaussian 
random field and are believed to be generated 
by quantum processes in an inflationary stage. 
We treat as example the possible presence of an intrinsic 
dipole component $(\frac{\delta T}{T})_{dipole} = 
\int n^i\frac{\delta T}{T}\frac{d\Omega}{4\pi}$ of the CMB, 
which is independent of the usual kinematic component 
due to observer motion relative to the CMB rest frame 
(Langlois 1966, Lineweaver 1966). In principle,
all terms in (19) may contain dipole contributions. The first 
term results from a spatial gradient of the radiation energy 
density along the past null geodesics and requires the use of 
perturbed geodesics to obtain a nonzero result. The second 
term is produced by large-scale 
gravitational potential variations and is calculated using 
the FRW approximation for the null geodesics. Its dipole 
contribution can be written
\beq
\left(\frac{\delta T}{T}\right)_k = 
\frac{1}{15} \int_{t_1}^{t_0} \frac{A_1}{a^2}(\dot h_{kll}
-\frac{1}{3}\dot h_{llk})
\eeq
with $A_1= \int_{t_1}^{t_0} dt/a$. 
\ni
The last term proportional to the acceleration of the fluid is
absent in the case of noninteracting dust. 

\bsk
\ni {\bf{7. CONCLUDING REMARK}}
\ssk
\ni
The MAP and PLANCK SURVEYOR satellite missions are expected to 
measure - together with other observations - many cosmological 
parameters such as the amplitudes and spectral indices of scalar 
and tensor fluctuations and the densities of various mass components 
of the universe. We emphasize that to these fundamental cosmological 
quantities {\it one has to add further geometrical and kinematical 
parameters} - the spatial curvature of the FRW models is only the 
simplest one. 

\ssk
\ni


\bsk
\baselineskip = 12pt
{\references \ni {\bf{REFERENCES}}
\ssk

\ref Bunn, E.F., Ferreira, P., Silk, J. 1996, Phys. Rev. Lett. 77, 2883
\ref Collins, C., Hawking, S. 1973, M.N.R.A.S. 162, 307
\ref Dunsby, P.K.S. 1997, Class. Quant. Grav. 14, 3391.
\ref Dunsby, P.K.S., Bruni, M., Ellis, G.F.R. 1992, ApJ 395, 54
\ref Ellis, G.F.R., Bruni, M. 1989, Phys. Rev. D 40, 1804
\ref Ellis, G.F.R., Hwang, J., Bruni, M. 1989, Phys. Rev. D 40, 1819
\ref Garcia-Bellido, J., Liddle, A.R., Lyth, D.H., Wand, D. 
     1995, Phys.Rev. D 52, 6750
\ref Grishchuk, L.P., Zeldovich, Ya.B. 1978, Astron. Zh. 55, 209
     [Sov. Astron. 1978 22, 125].
\ref Grishchuk, L.P. 1992, Phys. Rev. D 45, 4717
\ref Hawking, S. 1969, M.N.R.A.S. 142, 129
\ref Kashlinsky, A., Tkachev, I.I., Frieman, J. 1994, 
     Phys. Rev. Lett. 73, 1582
\ref Kristian, J., Sachs, R. K. 1966,  ApJ 143, 379
\ref Kogut, A., Hinshaw, G., Banday, A.J. 1997, 
     Phys.Rev. D 55, 1901
\ref Langlois, D. 1996 Phys. Rev. D 53, 2908, Phys. Rev. D 54, 2447,
     Helv. Phys. Acta 69, 229
\ref Lineweaver, C.H., Tenario, L., Smoot, G.F., Keegstra, P., 
     Banday, A.J., Lubin, P. 1996, astro-ph/9601151  
\ref Lyth, D.H. 1995, Ann. N.Y. Acad. Science 759, 701
\ref Misner, C.W., Thorne, K.S., Wheeler, J.A. 1973, Gravitation 
     (W.H.~Freeman and Company, San Francisco)
\ref Panek, M. 1986, Phys.Rev. D 34, 416
\ref Russ, H., Soffel, M., Xu, C.M., Dunsby, P.K.S. 1993, 
     Phys. Rev.,  D 48, 4552
\ref Turner, M.S. 1991, Phys. Rev., D 44, 3737

\end{document}